\documentclass[aps,prx,twocolumn,amsmath,amssymb,superscriptaddress,longbibliography]{revtex4-2}
\usepackage[breaklinks,colorlinks,bookmarks=false,citecolor=blue,linkcolor=red,urlcolor=blue]{hyperref}
\usepackage{blindtext,enumitem,graphicx,dcolumn,color,xcolor,amsmath,braket,bm,changes,chemformula,cleveref,times}

\graphicspath{{figures/}}
\usepackage{blindtext}
\usepackage{changes}

\definecolor{darkpastelgreen}{rgb}{0.01, 0.75, 0.24}

\crefname{equation}{Eq.}{Eqs.}
\Crefname{equation}{Eq.}{Eqs.}
\crefname{figure}{Fig.}{Figs.}
\Crefname{figure}{Fig.}{Figs.}
\crefname{section}{Sec.}{Secs.}
\Crefname{section}{Sec.}{Secs.}

\begin{document}
\title{Quantum Electrodynamics in 2+1 Dimensions as the Organizing Principle \\
of a Triangular Lattice Antiferromagnet}

\author{Alexander Wietek}
\email{awietek@pks.mpg.de}
\affiliation{Max Planck Institute for the Physics of Complex Systems, N\"othnitzer Strasse 38, Dresden 01187, Germany}
\affiliation{Center for Computational Quantum Physics, Flatiron Institute, 162 5th Avenue, New York, New York 10010, USA}

\author{Sylvain Capponi}
\affiliation{Laboratoire de Physique Th\'eorique, Universit\'e de Toulouse, CNRS, UPS, France}

\author{Andreas M. L\"{a}uchli}
\affiliation{Laboratory for Theoretical and Computational Physics, Paul Scherrer Institute, 5232 Villigen, Switzerland}
\affiliation{Institute of Physics, \'{E}cole Polytechnique F\'{e}d\'{e}rale de Lausanne (EPFL), 1015 Lausanne, Switzerland}

\begin{abstract}
Quantum electrodynamics in $2+1$ dimensions (QED$_3$) has been proposed as a critical field theory describing the low-energy effective theory of a putative algebraic Dirac spin liquid or of quantum phase transitions in two-dimensional frustrated magnets. We provide compelling evidence that the intricate spectrum of excitations of the elementary but strongly frustrated $J_1$-$J_2$ Heisenberg model on the triangular lattice is in one-to-one correspondence to a zoo of excitations from QED$_3$, in the quantum spin liquid regime. This includes a large manifold of explicitly constructed monopole and bilinear excitations of QED$_3$, which is thus shown to serve as an organizing principle of phases of matter in triangular lattice antiferromagnets and their low-lying excitations. Moreover, we observe signatures of emergent valence bond solid (VBS) correlations. This can be interpreted either as evidence of critical VBS fluctuations of an emergent Dirac spin liquid or as a transition from the $120^\circ$ N\'eel order to a VBS whose quantum critical point is described by QED$_3$. Our results are obtained by comparing ansatz wave functions from a parton construction to exact eigenstates obtained using large-scale exact diagonalization up to $N=48$ sites. 
\end{abstract}

\date{\today}

\maketitle

\section{Introduction}

\begin{figure}[t]
    \centering
    \includegraphics[width=\columnwidth]{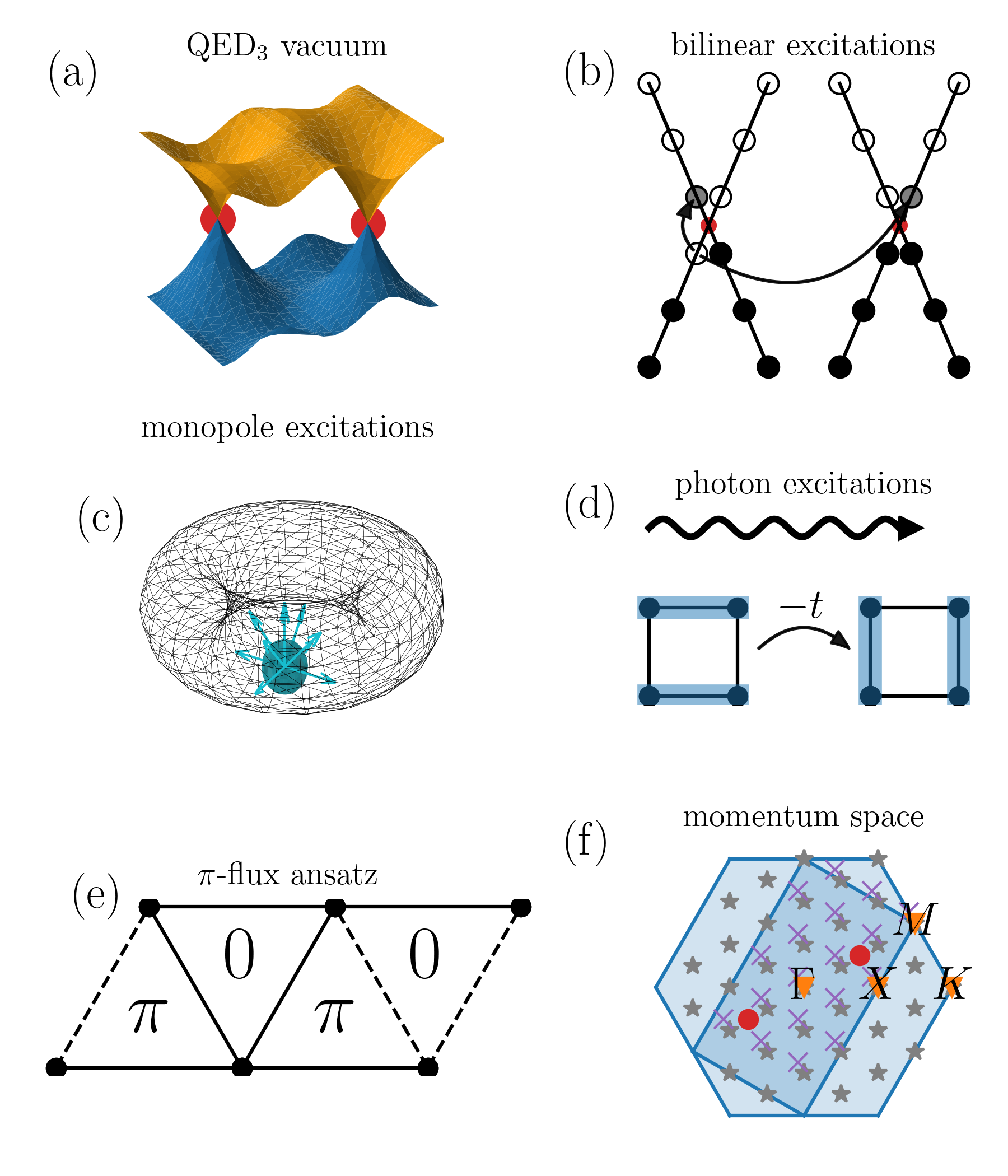}
    \caption{Quantum electrodynamics in $2+1$ dimensions implemented on the triangular lattice. (a) The QED$_3$ vacuum is constructed by filling the lower band of the $\pi$-flux ansatz shown in (e) and performing a Gutzwiller projection. The red dots indicate the Dirac nodes with linear dispersion. (b) Bilinear excitations are obtained by creating particle-hole excitations on the Dirac-sea. (c) Monopole excitations are created by distributing a unit of magnetic flux across the full torus. (d) Gauge field (photon) excitations are studied using an effective quantum dimer model. (e) The $\pi$-flux ansatz is implemented on a two-site unit cell and features $\pi$-flux through up-triangles and $0$-flux through down-triangles. An example gauge is shown with the dashed and solid lines which indicate a phase of $\pi$ and $0$ on the hopping, respectively. (f) Brillouin zone (BZ) of the triangular lattice and folded BZ of the $\pi$-flux ansatz. The high symmetry points $\Gamma, K, M, X$ are marked with orange symbols. The grey stars show the momentum resolution of the $N=36$ simulation cluster. The purple crosses indicate the shifted momenta minimizing the kinetic energy of the filled Dirac sea.}
    \label{fig:overview}
\end{figure}

The emergence of collective equations of motion from seemingly unrelated microscopic interactions is one of the most fascinating aspects of many-body physics. Strongly correlated electrons can realize intriguing quantum field theories (QFT) as their low-energy effective description which otherwise are used to describe the fundamental laws of elementary particles. Quantum spin liquids (QSL) in frustrated magnetism are a particularly exciting instance of emergent QFTs~\cite{Savary2016}. Topological QFTs describe certain gapped QSLs which have been shown both analytically and numerically to emerge in local spin models. This includes the emergent $\mathbb{Z}_2$ lattice gauge theory in the toric code or the Kitaev's honeycomb model~\cite{Kitaev2006} as well as Chern-Simons theories realized in chiral spin liquids~\cite{Kalmeyer1987,Wen1989}, which have been discovered in simple Heisenberg-like Hamiltonians on the triangular and kagome lattice~\cite{Bauer2014,He2014,Gong2014,Wietek2015,Wietek2017,Nataf2016,Szasz2020}.

The arguably most widely known QFT is quantum electrodynamics (QED). While on one hand, it is the fundamental theory of fermions coupled to a U($1$) gauge field describing the physics of elementary electrons and photons, it has also been discussed as an emergent field theory in frustrated magnets. Remarkably, QED in three spatial dimensions can be realized in pyrochlore spin ice compounds~\cite{Benton2012,Castelnovo2008,Castelnovo2012}. Condensed matter systems also allow for the realization of QED in less than 3 spatial dimensions. The physics of QED in $2+1$ dimensions (QED$_3$) is considered to be more strongly coupled than its $3+1$ dimensional counterpart while exhibiting a richer phenomenology than the confining $1+1$ dimensional QED, also referred to as the Schwinger model. However, the physics of QED$_3$ is until today still a subject of intense research.

Early on, QED$_3$ has been suggested as an effective field theory for so-called algebraic or Dirac spin liquids (DSL) in quantum magnets~\cite{Hastings2000,Hermele2004,Hermele2005}. While model wave functions of these states have been studied~\cite{Ran2007,Iqbal2013,Iqbal2016} and signatures of gapless spin liquids have been detected in certain spin models~\cite{He2017,Hu2019}, no realization of Dirac spin liquid has until today been unambiguously confirmed. However, several numerical studies have recently highlighted the relevance of QED$_3$ 
for elementary frustrated quantum magnets, such as the kagome Heisenberg antiferromagnet or the $J_1$-$J_2$ Heisenberg model on the triangular lattice.

In this work, we demonstrate that QED$_3$ serves as on organizing principle of the physics of triangular lattice Heisenberg antiferromagnets by exact numerical calculations. Specifically, we consider the spin-1/2 Heisenberg model on the triangular lattice, with nearest-neighbor $J_1$ and next-nearest-neighbor $J_2$ antiferromagnetic (AF) couplings,
\begin{equation}
\label{eq:model}
{\cal H} = J_1 \sum_{\langle ij\rangle} \mathbf{S}_i\cdot \mathbf{S}_j + J_2 \sum_{\langle\langle ij\rangle\rangle} \mathbf{S}_i\cdot \mathbf{S}_j.
\end{equation}
We construct a "vacuum" of QED$_3$ and a plethora of low-lying excitations directly on a lattice by means of Gutzwiller projection of a DSL parton ansatz. These model states of both the vacuum and the low-lying excitations are then directly compared to numerically exact ground state and low-energy eigenstates of an $N=36$ sites simulation cluster. Moreover, we present the complete low-energy spectrum with space group and spin-parity resolution for the $N=48$ sites cluster, obtained using large-scale Exact Diagonalization (ED). 
The $N=48$ ED spectrum used approximately 25 million CPU hours which was possible thanks to an overall 40 million CPU hours grant from PRACE.

The ground state phase diagram of \cref{eq:model} features a $120^\circ$ N\'eel ordered state for $J_2/J_1 \lesssim 0.09$~\cite{Huse1988,Bernu1994,Capriotti1999,White2007} and a stripy antiferromagnetic state for $J_2/J_1 \gtrsim 0.14$. In between, a paramagnetic regime is stabilized whose nature has been subject of intense debate~\cite{Kaneko2014,Zhu2015,Hu2015,Iqbal2016,Saadatmand2016,Oitmaa2020}. In particular, several references have pointed out the possibility of a Dirac spin liquid~\cite{Kaneko2014,Iqbal2016,Hu2019} or a gapped $\mathbf{Z}_2$ spin liquid~\cite{Zhu2015,Hu2015,Saadatmand2016,Jiang2022} being stabilized in this regime. Quite interestingly, the paramagnetic regime is highly sensitive to further perturbations: for instance, when adding a further neighbor AF coupling $J_3$, a gapless chiral spin liquid (CSL), which spontaneously breaks time and lattice reflection symmetry, was reported using DMRG~\cite{Gong2019}; similarly, adding an explicit tiny chiral term in the Hamiltonian ($J_\chi \bm{S}_i \cdot (\bm{S}_j \times \bm{S}_k)$) leads to a gapped CSL~\cite{Saadatmand2017,Gong2017,Wietek2017}.

\begin{figure*}[t]
    \centering
    \includegraphics[width=\textwidth]{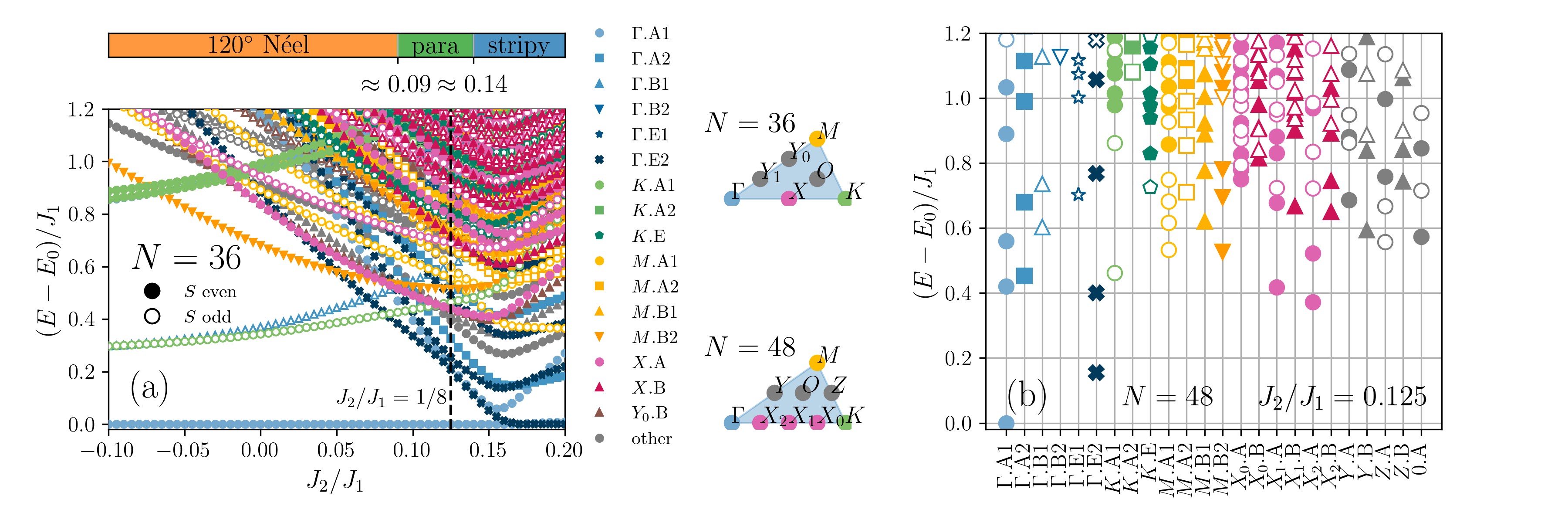}
    \caption{Energy spectrum of the $J_1$-$J_2$ model on the triangular lattice for $N=36$ as a function $J_2/J_1$ in (a) and for $N=48$ at $J_2/J_1=0.125$ in (b). Symbol colors and shapes indicate the momentum and point group representation of the eigenstate. Filled (open) symbols indicate even (odd) total spin $S$. The $120^\circ$ N\'eel state persists for $J_2/J_1 \lesssim 0.09$, the paramagnetic regime is realized for $0.09 \lesssim J_2/J_1 \lesssim 0.14$, and the stripy magnet is stabilized for $J_2/J_1 \gtrsim 0.14$. The vertical dashed line at $J_2/J_1=0.125$ in panel (a) highlights the spectrum in order to compare with the $N=48$ data shown in panel (b).}
    \label{fig:j1j2_spectrum}
\end{figure*} 

\section{QED$_3$ and its torus spectrum}

Based on a parton approach, it has been advocated that QED$_3$, i.e.~the quantum field theory of $N_f$ flavors of massless Dirac fermions coupled to a compact $U(1)$ gauge field in 2+1D, could be an appropriate description of the low-energy physics of certain spin liquid phases, in particular for the kagome and the triangular lattice Heisenberg models. In these examples, the number of fermion flavors is $N_f=4$.

In our work, we want to study the low-energy spectrum of finite-size spin systems and investigate whether the spectrum is in agreement with the low-energy spectrum of QED$_3$ in finite volume. In order to execute this program we need to review what is known about QED$_3$ and how the finite size low-energy spectrum of a quantum field theory on a torus is structured.

It is useful to start understanding QED$_3$ in the limit of $N_f\rightarrow \infty$. In that limit, the gauge fluctuations are suppressed and the theory has a simpler structure. The theory is a conformal field theory (CFT) with a variety of gapless modes. On the one hand, the bare massless fermions remain gapless, as well as the gauge field in the form of photons. Another set of excitations, the monopoles -- which will play an important role later --, are found at a large energy scale which is proportional to $N_f$ in the limit $N_f\rightarrow\infty$. In the correspondence to the spin model, only the gauge invariant and charge neutral excitations of QED$_3$ are allowed to appear, therefore only neutral fermion excitations are visible (e.g.~vacuum, neutral bilinears c.f.~Fig.\ref{fig:overview}(b)) in addition to the gauge field excitations and the monopoles, see Fig.\ref{fig:overview}(c). As $N_f$ decreases towards the value of interest $N_f=4$, the monopoles also become gapless low-energy excitations. The role of the monopoles is also crucial for the stability of the CFT window of QED$_3$ itself. In the extreme limit $N_f=0$ it is known since Polyakov~\cite{Polyakov1977} that a pure compact $U(1)$ gauge theory in 2+1D is confining in the presence of a UV cutoff. Analytical and numerical results on the $N_f$ extent of the conformal window of QED$_3$ did not yet reach a consensus, but it is conceivable that $N_f=4$ still belongs to the conformal window~\cite{Karthik2016,DiPietro2017,Xu2019,Albayrak2022}.

As QED$_3$ is a CFT for large enough values of $N_f$, one can characterize the low energy excitations based on their scaling dimensions. The current best estimates for the scaling dimensions at $N_f=4$ are summarized in Ref.~\cite{Albayrak2022}. For 1+1D systems described by a CFT, the energy spectrum on a circle is equivalent to the operator content on the 2D (space-time) torus and harbors thus the spectrum of scaling operators and their descendants, arranged into Verma modules. In 2+1D CFTs an analogous correspondence only holds for Hamiltonians quantized on a spatial {\em sphere}, while the lattice models studied in condensed matter physics more naturally live on a spatial {\em torus}. Recently the torus energy spectrum for a series CFTs, such as the Wilson-Fisher and Gross-Neveu-Yukawa theories have been studied using a combination of numerical and analytical 
results~\cite{Schuler2016,Whitsitt2016,Thomson2017,Whitsitt2017,Schuler2021}. The spectrum is understood to collapse as $1/L$ ($L$ being the linear extent of the system) as expected for a relativistic theory. The low-energy spectrum $\Delta_i=E_i-E_\mathrm{GS}$ multiplied by $L$ then forms a fingerprint $\xi_i\equiv \Delta_i \times L$ of the conformal field theory governing the low-energy spectrum~\footnote{We set the effective speed of light $v=1$ here for simplicity.}.  While the torus spectrum of these theories is known to be different from the sphere spectrum~\cite{Belin2018}, in the above works a phenomenological reminiscence among some of the low-lying states on the sphere and on the torus has been observed. So we expect the low-energy part of the torus spectrum of QED$_3$ to be formed by the torus doppelg\"anger of the vacuum, the neutral bilinears, the monopoles, and gauge-field excitations. 

The torus spectrum of QED$_3$ on a (square) torus has been studied in the limit $N_f\rightarrow\infty$ in Ref.~\cite{Thomson2017}. This includes the bilinears and the photon excitations. We discuss the basic structure of the neutral fermionic sector adapted to the hexagonal Brillouin zone torus in App.~\ref{app:bilinear}. A hallmark of the fermionic bilinear spectrum is the massive degeneracy of levels already at the first fermionic bilinear excitation above the ground state (larger than their number $N_f^2$ on the sphere), combined with a rather soft photon excitation which additionally boosts the number of low-lying levels. So even in the absence of monopoles, the low-energy torus spectrum of QED$_3$ is much denser than for example the torus spectrum of a 2+1D Ising CFT~\cite{Schuler2016}.

The monopole excitations in lattice systems carry quantum numbers of the space and the spin symmetry group. In Refs.~\cite{Song2019,Song2020} the quantum numbers of the first $q=\pm 1$ monopoles for $N_f=4$ have been studied for various lattices. For the triangular lattice of interest here, spin-singlet monopoles at each of the six $X$ points in the Brillouin zone have been identified together with spin-triplet monopoles at each of the two $K$ points. Altogether these form twelve monopole states in total. We expect precisely these states to be present at low energy in the torus spectrum. Note that the $SU(4)$ symmetry of QED$_3$ for $N_f=4$ predicts the energies of the 12 monopole states to be degenerate, also on the torus. Since this symmetry is not exact on the lattice but only emerges in the IR, one would expect some finite-size splitting to lift the degeneracy.

In the absence of an analytical result for the torus spectrum for $N_f=4$ we proceed now by a Gutzwiller projection procedure to generate model state wave functions for several classes of torus excited states of QED$_3$ on the triangular lattice. This way we will be able to describe the vacuum, the bilinears, and the monopoles, see Sec.~\ref{sec:qed3_to_spinmodel}. While we cannot predict the $\xi_i$ values of the torus spectrum this way, we can still check whether the significant overlaps of the Gutzwiller projected states cover relevant low-energy states of the ED spectrum on a given system size. In the next step, we can then interpret the relative excitation energies of the different classes of excitations, and possibly gain some insights about the relevance of QED$_3$ for the low energy physics of the triangular lattice $J_1-J_2$ Heisenberg model, and perhaps even about the stability of QED$_3$ itself. 

Note that we are not able to generate excitations of the gauge field sector in this way. In order to shed some light on this sector, we take a different approach and consider the spectrum of the Rokhsar-Kivelson quantum dimer model~\cite{Rokhsar1988} on the same finite-size clusters, see Sec.~\ref{subsec:gaugefluctuations}, for comparison.

\section{Spectrum of the $J_1$-$J_2$ triangular antiferromagnet}

We now turn to the main problem of interest, to understand the low-energy spectrum of the $J_1-J_2$ Heisenberg model on the triangular lattice and the nature of the corresponding phases. The low-energy spectrum in many phases of matter is well understood, for example in magnetically ordered phases~\cite{Wietek2017a}, or in topologically ordered phases with their torus ground state degeneracy. As discussed in the previous section, torus spectra for conformal field theories in 2+1D are however a topic of ongoing research.

We study the energy spectra of \cref{eq:model} obtained from large-scale exact diagonalization~\cite{Wietek2018} where we resolve the spectrum w.r.t.~the irreducible representations (irreps) of translational, point group (PG), and spin-flip symmetries in \cref{fig:j1j2_spectrum}. States in the even (odd) spin-flip representation have even (odd) total spin $S$. We denote the irrep of these states by $\bm{k}$.$\rho$, where $\bm{k}$ labels the momentum and $\rho$ labels the point group irrep.  
For $N=36$ \cref{fig:j1j2_spectrum}(a) we cover the range of $J_2/J_1 \in [-0.1, 0.2]$. The ground state for $J_2/J_1 \lesssim 0.09$ is magnetically ordered with a $120^\circ$ N\'eel pattern and ordered with a stripy pattern for $J_2/J_1 \gtrsim 0.14$~\cite{Lecheminant1995}. In between, a quantum paramagnetic regime is stabilized, whose nature is the main subject of our discussion. Far outside the paramagnetic regime, e.g.~for $J_2/J_1=-0.1$ in the $120^\circ$ N\'eel phase or at $J_2/J_1=0.2$ in the collinear stripy phase, the main structure of the low-lying energy spectrum can be well understood by tower-of-states (TOS) analysis~\cite{Lecheminant1995,Wietek2017a}. In the $120^\circ$ N\'eel phase, TOS analysis predicts an $S=0$ ground state in the $\Gamma$.A1 sector whereas the first two quasi-degenerate excited states are in the $K$.A1 and $\Gamma$.B1 sector with $S=1$, precisely what is observed in the $J_1$-$J_2$ model in \cref{fig:j1j2_spectrum}(a). Similarly, TOS analysis in the stripy phase predicts three quasi-degenerate $S=0$ states, one at $\Gamma$.A1 and two forming a $\Gamma$.E2 irrep. For details on the TOS analysis and further predictions see App.~\ref{app:tos} and Ref.~\cite{Wietek2017a}. In the intermediate regime, however, the spectrum is rather dense and low-lying excitations belong to various irreps of the symmetry group. As shown in the subsequent sections, many of these levels can be identified with non-trivial excitations on top of the QED$_3$ vacuum. In particular, the prominent singlet levels with momentum $\mathbf{k}= X$ and $\mathbf{k}= M$, which are neither part of the TOS for the $120^\circ$ N\'eel nor stripy order, will be related to monopole and bilinear excitations of QED$_3$. \cref{fig:j1j2_spectrum}(b) shows the low-energy spectrum of the $N=48$ cluster at $J_2/J_1=0.125$ in the paramagnetic regime. Again, we observe several low-lying singlet excitations (filled symbols) at non-trivial momenta and PG irreps. In particular, two singlet $\Gamma$.E2, one $\Gamma$.A2, and two levels at $X_1$.A and $X_2$.A which are sixfold degenerate can be found below the spin gap, and lower than their $N=36$ occurrence. The lowest lying triplets can be found in the $K$.A1, $Z$.A and $M$.A1 sectors, i.e.~along the Brillouin zone boundary at roughly the same energy. We refer to App.~\ref{app:spectrum_36_48} for a detailed comparison of the two system sizes studied here. 

\begin{figure*}
    \centering
    \includegraphics[width=\textwidth]{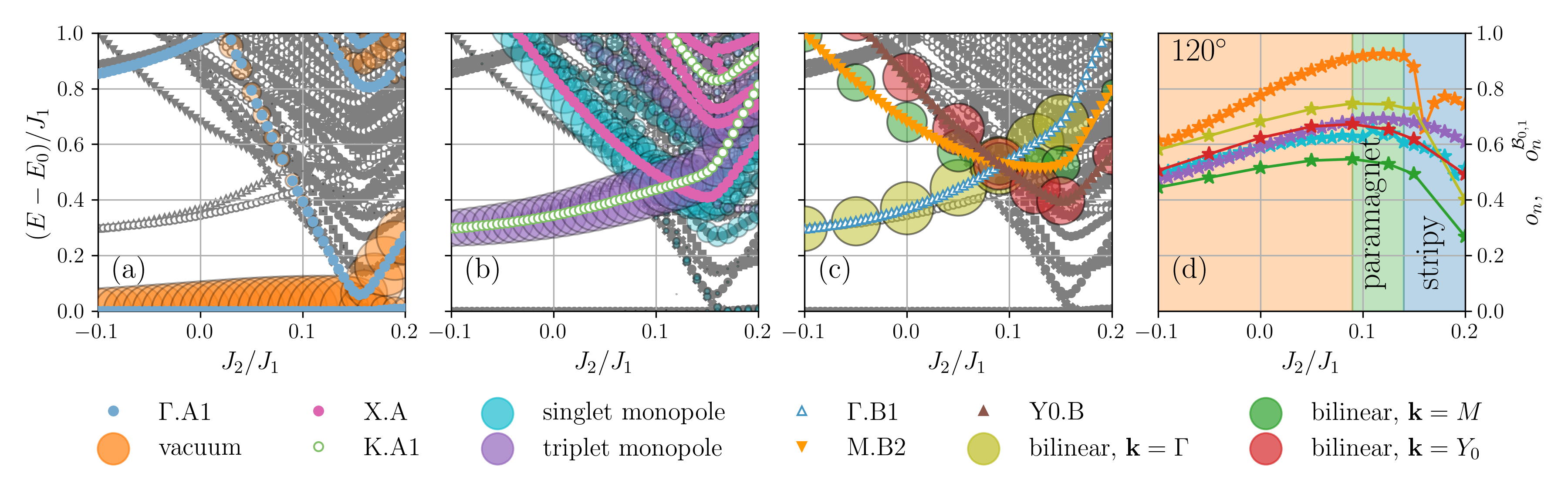}
    \caption{Overlaps of low-energy levels of the $J_1$-$J_2$ model with various ansatz wave functions constructed out of various excitations above the QED$_3$ vacuum. For the vacuum and monopole states, we show $o_n$, whereas for the bilinear excitations $o_n^{\mathcal{B}_{0,1}}$ is shown. The diameter of the colored circles is proportional to the overlap with the state at the center of the circle. (a) The vacuum state has significant overlap (up to $\approx 0.92$) with the ground state.  (b) The singlet monopole has significant overlap with the low-lying $X.A$ level ($\approx 0.65$) while the triplet monopole has sizeable overlap with the low-lying triplet $K.A1$ ($\approx 0.69$) level. (c) Among the  bilinear-excitations $\mathcal{B}_0$ there is significant overlap with the low-lying $\Gamma.B1$ and $M.B2$ levels. The $Y_0.B$ energy level has significant overlaps ($\approx 0.67$) with the bilinear excitations $\mathcal{B}_1$ (d) numerical values of the maximal overlap of the aforementioned states with eigenstates from ED. For every excitation, the maximum is attained in the paramagnetic regime.}
    \label{fig:overlaps}
\end{figure*}

We will provide an attempt at understanding the presence of these low-energy states from the QED$_3$ perspective in the following.

\section{From QED$_3$ to the spectrum of the $J_1$-$J_2$ model}
\label{sec:qed3_to_spinmodel}
To elucidate the structure of the energy spectrum in the intermediate paramagnetic regime, we will now relate them to  systematically constructed wave functions of excitations of the $\pi$-flux ansatz (cf, \cref{fig:overview}(a, e)) for the Dirac spin liquid. Our figure of merit to compare ansatz wave functions $| \psi_{GW} \rangle$ with exact eigenstates $|\psi_n\rangle $ of  \cref{eq:model} is the overlap 
\begin{equation}
    o_n \equiv |\langle \psi_{GW} | \psi_n \rangle|.
\end{equation}
Note that $o_n$ is expected to approach zero exponentially fast as $N\rightarrow \infty$ due to the orthogonality catastrophe. But on a given system size the behavior of $o_n$ as a function of $J_2/J_1$ is insightful. The overlap calculations have been performed in the full Hilbert space without employing any symmetry.

\subsection{Parton construction}
Ansatz wave functions of QSLs can be systematically constructed using the parton construction~\cite{Wen2002,Wen2007}. The original spin operator $\bm{S}_i$ is decomposed into fermionic parton operators $f_{i\alpha}$, ($\alpha=\uparrow,\downarrow$) acting on an enlarged fermionic Hilbert space, 
\begin{equation}
    \label{eq:partons}
    \bm{S}_i = \frac{1}{2}f_{i\alpha}^\dagger \bm{\sigma}_{\alpha\beta} f_{i\beta},
\end{equation}
where $\bm{\sigma} = (\sigma_x, \sigma_y, \sigma_z)$ denote the Pauli matrices. To impose the single occupancy constraint $n_{i\alpha} = f_{i\alpha}^\dagger f_{i\alpha}=1$, the partons can be coupled to a dynamical gauge field~\cite{Wen2007}. Rewriting a spin model such as \cref{eq:model} into fermionic operators yields a Hamiltonian with quartic interactions in the parton operators which are coupled to a dynamical gauge field. Applying a Hubbard-Stratonovich transformation and assuming a static gauge field, the resulting parton Hamiltonian is simplified into a mean-field model of the form,
\begin{equation}
    \label{eq:partonmeanfield}
    H_{mf} = \sum_{i,j,\alpha} \chi_{i,j} f^\dagger_{i,\alpha}f_{j,\alpha} + \rm{h.c.},
\end{equation}
where $\chi_{i,j}$ denote the hopping amplitudes of the particular mean-field ansatz. The $\pi$-flux ansatz of the DSL on the triangular lattice we employ is shown in \cref{fig:overview}(e)~\cite{Iqbal2016,Lu2016}. A magnetic flux of $\pi$ is implemented through the "up" triangles while zero flux is chosen through the "down" triangles. The bandstructure is shown in \cref{fig:overview}(a) and features two gapless Dirac cones with linear dispersion, cf. App.~\ref{app:dispersion}. To explicitly construct numerical ansatz wave functions $| \psi \rangle$, we perform a Gutzwiller projection of parton Slater determinants $| \psi_{\mathrm{free}} \rangle$,
\begin{equation}
    \label{eq:gutzwiller}
    | \psi_{GW} \rangle = \mathcal{P}_{GW}| \psi_{\mathrm{free}} \rangle = \prod_{i} (1 - n_{i\uparrow}n_{i\downarrow}) | \psi_{\mathrm{free}} \rangle.
\end{equation}
The Gutzwiller projection operator $ \mathcal{P}_{GW} = \prod_{i} (1 - n_{i\uparrow}n_{i\downarrow})$ imposes exactly the single occupancy constraint. 

Past studies focused mostly on variational studies based on the Gutzwiller vacuum, but a systematic and exhaustive study of Gutzwiller projected excited states 
has not been attempted before. In Ref.~\cite{Ferrari2019} a similar basis of Gutzwiller projected excited states was used to model dynamical response functions
by projecting the Hamiltonian into the Gutzwiller excitation subspace, while we investigate the overlap of the Gutzwiller subspace with the {\em exact} eigenstates. 

\subsection{Vacuum as the filled Dirac sea}
To construct a canonical ground state ansatz, we fill all single-particle energy levels of the lower band and perform a Gutzwiller projection. In our ansatz, the boundary conditions can be twisted, leading to a shift of the momentum grid in reciprocal space. The twist corresponds to changing the ansatz for the gauge field (i.e. the hopping phases) in mean-field Hamiltonian \cref{eq:partonmeanfield}~\footnote{This operation changes the flux through lines wrapping around the torus, while the flux $\pi$ or $0$ across elementary triangles is unaffected.}. We find that the energy of the mean-field ansatz is minimized whenever the two Dirac nodes, indicated as red dots in \cref{fig:overview}(f), are shifted to the center of three resolved momenta of the finite size cluster. We refer to this choice as the \textit{centered} boundary conditions, cf.~App.~\ref{app:centered_bv} 
The resolved single particle momenta on the $N=36$ site cluster with centered boundary conditions are shown as purple crosses in \cref{fig:overview}(f). We find that, also after Gutzwiller projection, the state with centered boundary conditions has lower variational energies than the standard periodic boundary conditions. Since our original model is real, we only consider the real part of the wave function after Gutzwiller projection (the centered Gutzwiller wave function is genuinely complex, while for the more common periodic/antiperiodic choices it would be real),
\begin{equation}
    |\psi_{vac}\rangle = \rm{Re} \left[ \mathcal{P}_{GW}| \psi_{D.S.}^{center} \rangle \right],
\end{equation}
where $|\rm \psi_{D.S.}^{center} \rangle$ denotes the filled Dirac sea in the centered gauge~\footnote{Note that we only take the real part of the Gutzwiller projected wavefunction for the vacuum, while all excited states overlaps are performed with the complex wavefunctions.}.

We computed overlaps of $|\psi_{vac}\rangle$ with low-lying energy states~\cite{Iqbal2021kagome} using a Krylov technique which avoids calculating all low-lying eigenfunctions explicitly~\cite{Gagliano1987}, see \cref{app:overlaps} for details. \cref{fig:overlaps}(a,d) show substantial overlaps with the ED ground state of up to $o_0=0.923$ at $J_2/J_1=0.12$. This maximum is attained in the paramagnetic regime and decays in the $120^\circ$ N\'eel ordered and stripy phase. In the stripy phase, the state with maximal overlap is at low energy but distinct from the ground state.

\subsection{Monopoles}

In the version of QED$_3$ we consider, the $U(1)$ gauge field is compact and as a consequence, there exist charge neutral, topological excitations known as monopoles. If these defects proliferate, the DSL will become unstable and will enter a confined phase, in our context e.g.~the familiar 120$^\circ$ magnetic N\'eel order (for triplet monopoles) or a 12-site unit cell valence-bond solid (VBS) that breaks lattice symmetries (for singlet monopoles)~\cite{Song2019}. Conversely, in the limit of a large number of flavors $N_f$, monopoles will be present but are irrelevant for the low energy physics. Hence, their precise role for a given $N_f$ is quite topical and still unclear.

Using approximately known scaling dimensions, on the triangular lattice the only symmetry-allowed three-monopole operator should be irrelevant~\cite{Song2019,Albayrak2022}. The quantum numbers of the single monopole operators on the torus are nontrivial and have been obtained in~\cite{Song2019,Song2020}: the singlet monopoles are in the $X$.A irrep while the triplet monopoles have momentum $K$ and are even under reflection and rotation (A1). 

Hence, a possible signature of the triplet monopole would be a low-energy mode at momentum $K$ as advocated in some numerics~\cite{Hu2019,Ferrari2019,Sherman2022,Drescher2022,Tang2022} Note however that this "exotic" monopole is evolving into a TOS level in the 120$^\circ$ phase, and therefore its presence at low energy is not as much as a surprise as for the singlet. In this work, we want to be more precise and we specifically construct a microscopic monopole wave function by fixing a flux pattern such that a $\pm 2\pi$ flux is distributed over the whole lattice (i.e. each triangular plaquette has an additional flux $\pm 2\pi/(2N)$ w.r.t. to the $\pi$-flux Ansatz), see App.~\ref{app:monopole}. By computing the tight-binding dispersion, we find an exact two-fold degenerate zero-energy state. Thus, we can generate several monopole wave functions by filling up all negative energy states and putting 2 electrons (up or down) in these levels, and then Gutzwiller projecting them. In the end, we can build six singlet and two triplet monopole wave functions from which we can compute exact overlaps with the many-body eigenstates. 

Results are presented in Figs.~\ref{fig:overlaps}(b,d) for the triplet and singlet monopole states. Quite remarkably, we do measure a significant overlap with the lowest singlet state found in the irrep $X$.A and the lowest triplet in the irrep $K$.A1, in perfect agreement with the expected quantum numbers. 
Both monopole excitations maximize $o_n$ in the paramagnetic regime, with overlaps of up to $o_n \approx 0.65$ for the singlet and  $o_n \approx 0.67$ for the triplet monopole, Fig.~\ref{fig:overlaps}(d). Moreover, we have also computed corresponding overlaps for two-monopole states, which are discussed in \cref{app:twomonopoles}.

\subsection{Bilinear fermionic excitations}

\begin{figure}
    \centering
    \includegraphics[width=\columnwidth]{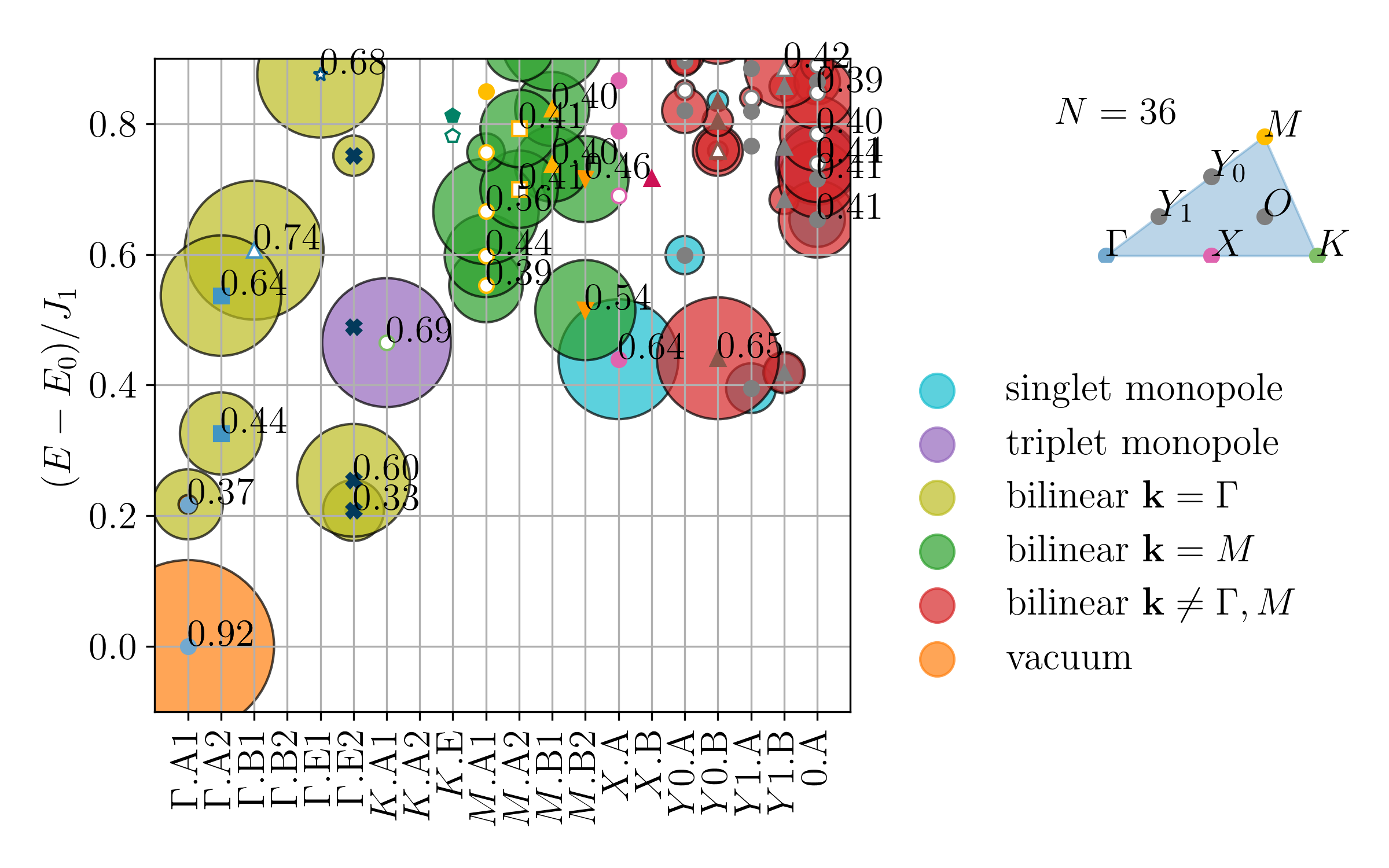}
    \caption{Overlaps of low-energy levels at $J_2/J_1 = 0.125$ with the the QED$_3$ ansatz wave functions from Gutzwiller projection.
    For the vacuum and monopole states, we show $o_n$, whereas for the bilinear excitations $o_n^{\mathcal{B}_{0,1}}$ is shown.
    The diameter of the colored circles is proportional to the overlap with the state at the center of the circle. We observe that almost every eigenstate in the dense low-energy ED spectrum has significant overlap with only one of the various excitation ansatz types. 
}
    \label{fig:spectrumexhaustion}
\end{figure}

The next set of excitations of QED$_3$ we investigate is particle-hole excitations of the Dirac sea represented as bilinears in the fermionic operators. 
Here, we construct them as particle-hole excitations of the parton ansatz, as shown in Fig.~\ref{fig:overview}(b) with centered boundary conditions. First, we consider excitations of partons in the Dirac sea at single particle momenta closest to the Dirac node. For simplicity, we focus on $S_z=0$ wave functions at momenta $\Gamma$ and $M$, i.e.~we can construct $24$ states which are all found to be linearly independent after Gutzwiller projection. These excitations are constructed by exciting a parton of either spin $\sigma = \uparrow,\downarrow$ on one of the six momenta closest to the two Dirac nodes, either to the exact same momentum (resulting in a $\bm{k}= \Gamma$ or 
one of the $\bm{k}= M$ point excitation) or the corresponding momentum at the other Dirac node (resulting in the other two $\bm{k}= M$ excitations). App.~\ref{app:bilinear} explains the counting of the degeneracies of the bilinear excitations in further details.

These $24$ states span a large subspace which we compare to low-lying excitations of the spin Hamiltonian. To do so, we orthogonalize the states after Gutzwiller projection such that we can compute the overlap of states from ED with this $24$ dimensional subspace we name $\mathcal{B}_0$. To compare ED eigenstates to a subspace $\mathcal{B}$, we define the overlaps $o_n^{\mathcal{B}}$, 
\begin{equation}
    \left(o_n^{\mathcal{B}}\right)^2 \equiv 
    \parallel \mathcal{P}_{\cal B} \ket{\psi_n} \parallel^2 = 
    \sum_{\alpha=1}^{\textrm{dim}(\mathcal{B})} |\langle \phi_\alpha | \psi_n \rangle|^2  ,
\end{equation}
where $|\phi_\alpha \rangle$ denotes an orthonormal basis of $\mathcal{B}$ and $\mathcal{P}_{\cal B} = \sum_{\alpha=1}^{\dim{\cal{B}}}\ket{\phi_\alpha}\bra{\phi_\alpha}$ denotes the projector onto the space $\cal B$.

Our definition is equivalent to measuring the norm of the projected state onto the space $\cal B$, since
\begin{equation}
(o_n^{\cal B})^2 = 
\parallel \sum_{\alpha=1}^{\dim{\cal{B}}}\ket{\phi_\alpha}\braket{\phi_\alpha | \psi_n} \parallel^2=\sum_{\alpha=1}^{\dim{\cal{B}}}|\braket{\phi_\alpha | \psi_n}|^2
\end{equation}
which yields $o_n^{\cal B} = 1$ if $\psi_n$ is an element of the space $\cal{B}$, and $o_n^{\cal B} = 0$ if the state is orthogonal. 

By construction, non-zero overlaps are only found with states at momenta $\Gamma$ or $M$ due to the momentum conservation of the Gutzwiller projection. An overview of all states with significant overlap is shown as yellow or green symbols in Fig.~\ref{fig:spectrumexhaustion}. Prominently, the bilinear excitations have a sizeable overlap of up to $o_n^{\mathcal{B}_0} \approx 0.74$ with the low lying $S=1$ $\Gamma.B1$ level belonging to the tower of states of the $120^\circ$ N\'eel state along with the $S=1$ $K.A1$ level, which we previously found to have large overlaps with the triplet monopole excitation. Interestingly, also the lowest lying $S=0$ level at $M.B2$ has sizeable overlaps of up to $o_n^{\mathcal{B}_0} \approx 0.54$. Regarding the point-group quantum numbers, some quantitative predictions have been made in Ref.~\cite{Song2019}: the bilinear excitations are expected at $\Gamma$ or $M$ and they are all odd under reflection. We do confirm large overlaps with odd states ($\Gamma$.E2 or $M$.B2), but we also find states which are even under reflection. As Ref.~\cite{Song2019} considered the $N_f^2=16$ neutral bilinear field operators, the torus geometry with its more complex excitation spectrum explains this discrepancy in the number of bilinear states, as emphasized in Ref.~\cite{Thomson2017}.

Furthermore, we have considered bilinear excitations with the minimal momentum transfer distinct from $\Gamma$ and $M$, which are degenerate in energy before projection with the already considered set. This yields another set of $48$ $S_z = 0$ ansatz wave functions. Again, we find these states to be linearly independent, spanning a space we call $\mathcal{B}_1$. Many low-lying exact eigenstates are found to have significant overlaps $o_n^{\mathcal{B}_1}$ with this space, cf.~\cref{fig:spectrumexhaustion}. Most notably, the low-lying $Y_0.B$ state has overlaps of up to $o_n^{\mathcal{B}_1} \approx 0.67$. We would like to emphasize the large number of low-lying bilinear excitations, $24\; (\bm{k}=\Gamma, M) + 48 \;(\bm{k}\neq \Gamma, M)=72$ states with $S_z=0$ and another $72$ states with total $S_z=\pm1$ expected for the DSL on the torus. Thus, there are a total of $144$ low-lying bilinear excitations of the parton ansatz, cf. \cref{app:bilinear}. The described  counting is valid, whenever each Dirac point has three symmetric neighboring momenta, whose quasiparticle energy bands have exactly the same energy. This holds for both the $N=36$ and $N=48$ site clusters in this manuscript as well as more generically $6N \times 6N$ clusters for $N\geq 1$.

As a remarkable result of this construction, we find that almost the entire complex and dense low-energy spectrum in the paramagnetic regime has significant overlaps, with either the vacuum state, monopoles or bilinear excitations, as is beautifully visible in~\cref{fig:spectrumexhaustion}. Again, the overlaps of the bilinear excitations are maximized in the paramagnetic regime, while sharply dropping off in the stripy phase. Quite interestingly, even in the $120^\circ$ N\'eel phase, some overlaps are still significant.
We interpret this as a sign that the $120^\circ$ N\'eel is a natural descendant of the DSL, while the collinear stripy phase has not been identified as an instability of the
DSL.

\subsection{Quantum dimer model}
\label{subsec:gaugefluctuations}

While we have a way to construct "model states" for the vacuum, the monopoles, and the fermion bilinear excitations, we are not aware of a correspondingly simple way to construct the gauge field states of QED$_3$ on the torus. At $N_f\rightarrow\infty$ the photon modes have been calculated for the square torus~\cite{Thomson2017}, and the softest gauge field excitation lies even below the first bilinear level on the square torus. At $N_f=4$ there are no corresponding results available.

We, therefore, pursue a rather different avenue here, by investigating the energy spectrum of the hardcore 
quantum dimer model (QDM) on the triangular lattice and comparing it to the $J_1-J_2$ Heisenberg model on the same finite-size clusters. There are two different points of view on this procedure. The first one is that there is a long history in frustrated quantum magnetism to investigate QDMs as effective models for the singlet subspace of $S=1/2$ Heisenberg Hamiltonians in magnetically disordered phases, such as VBS or spin liquid phases~\cite{Rokhsar1988,Mambrini2000,Poilblanc2010,Rousochatzakis2014,Mambrini2015}. Most of these applications were for square and kagome lattices, but none for the triangular lattice so far \footnote{Note that the triangular lattice QDM is well studied, but not with the purpose of the specific modeling of a spin-disordered Heisenberg model}. The other point of view is to consider the QDM as a quantum link model for a pure gauge theory possibly with static background charges~\cite{Wiese2021}. In this context however, the triangular lattice QDM would be expected to describe a $\mathbb{Z}_2$ gauge theory and not a $U(1)$ theory. The tension between these two points of view needs to be clarified in future work in view of our findings below.

The triangular lattice QDM was shown \cite{Moessner2001} to host an intriguing $\mathbb{Z}_2$ spin liquid for values of $V/t \in [0.8,1]$, i.e~in the vicinity of the Rokhsar-Kivelson point $V/t=1$. For smaller $V/t$ values a valence bond solid with a 12-site unit cell is realized with an estimated parameter extent in $V/t \in [-0.75\pm0.25,0.8]$~\cite{Moessner2001, Ralko2005}. Our own exact diagonalization spectra shown in \cref{fig:qdm_spectrum} suggest that $V/t = -1.0$ is still within the VBS phase for the considered clusters. For even more negative values of $V/t$ a columnar VBS phase is found. 

\begin{figure}
    \centering
    \includegraphics[width=\columnwidth]{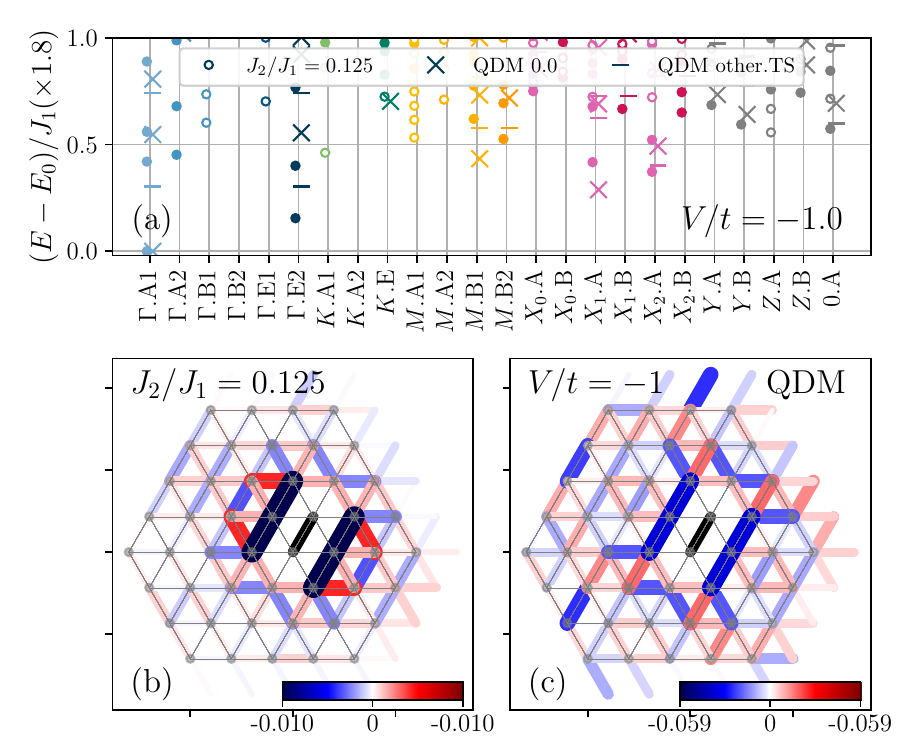}
    \caption{
    (a) Comparison of the low-lying energy spectrum of the $J_1$-$J_2$ model at $J_2 / J_1 = 0.125$ and the QDM model at $V/t=-1.0$ on the $N=48$ cluster. The filled (open) circles denote even (odd) spin levels. The QDM is only expected to reproduce filled levels because it cannot describe states with total spin $S>0$.
    (b) Connected dimer correlations $\langle(S_0^z S_1^z)(S_i^z S_j^z)\rangle_c$ of the ground state from ED of the $J_1$-$J_2$ model at $J_2 / J_1 = 0.125$ on the $N=48$ cluster. (c) Connected dimer correlations of the QDM in the VBS phase at $V/t=-1.0$. We observe a close resemblance between these patterns. }
    \label{fig:vbsqdm}
\end{figure}

Interestingly we find that many singlet levels of the low energy spectrum of the Heisenberg model at $J_2/J_1=0.125$ are surprisingly well reproduced by the low-energy spectrum of the QDM at $V/t=-1.0$, see Fig.~\ref{fig:vbsqdm}(a). We observe that the first excited state in both cases is a singlet at $\Gamma.E2$. The next low-lying excitations are two excitations at $X_1.A$ and $X_2.B$ for both the QDM and the spin model. These excitations are then followed by two singlet excitations at $M.B1$ and $M.B2$. Moreover, we observe a low-lying singlet state at $K.E$, which cannot be described as a monopole or bilinear excitation, see Fig.~\ref{fig:spectrumexhaustion}. These momenta are exactly what is expected for a lozenge VBS, cf. \cref{tab:andersontowermagtri}. However, we could not match all of the respective point group irreps, which can differ in a spin model, where non-trivial phases of the resonating dimers can occur. This unexpected (but not perfect) similarity of the low-energy spectrum of the two models raises the possibility that the QED$_3$ region of the triangular lattice is actually unstable and flows to a confining phase in the IR. The known QDM phase diagram suggests that this confining phase could be a 12-site unit cell VBS. In order to probe for this possibility we have calculated the dimer-dimer correlations of both models in Fig.~\ref{fig:vbsqdm}(b,c). While the qualitative agreement between the correlations of the two models is remarkable, in the spin model the correlations decay faster with distance~\cite{Roth2023}. It will be interesting for future research to explore whether there is a small but finite VBS order parameter in the spin liquid region, or whether these correlations are ultimately just the critical VBS fluctuations expected in the QED$_3$ Dirac spin liquid.

\section{Discussion / Conclusion}

In the above, we revealed a rather compelling one-to-one correspondence between the elementary excitations of QED$_3$ and the excitations of the triangular $J_1$-$J_2$ Heisenberg model, see Fig.~\ref{fig:spectrumexhaustion}. Even though the structure of low-energy excitations in the paramagnetic regime is rather complex, we have demonstrated that monopole and bilinear excitations of the $\pi$-flux ansatz have comprehensive overlaps with almost all low-lying eigenstates. From this, we can compellingly conclude QED$_3$ to be the organizing principle of the phases of matter in and close to the paramagnetic regime. Moreover, we have pointed out a close resemblance between the dimer correlations and low-lying energy spectrum of the $J_1$-$J_2$ model in the paramagnetic regime and the quantum dimer model in the valence bond solid phase on both the $N=36$ and $N=48$ site clusters. This could constitute evidence for a valence bond solid being realized, also found in recent large-scale DMRG simulations pointing out quasi-long range dimer order~\cite{Jiang2022}. In light of our findings, we would like to discuss possible scenarios for the phase diagram of the $J_1$-$J_2$ model. 

A transition from the $120^\circ$ N\'eel phase to the $12$-site valence bond solid appears to be a possible scenario. From a field theoretical perspective, such a transition can be described by a deconfined quantum critical point~\cite{Senthil2004,Senthil2004b,Shackleton2021}. As pointed out in Ref.~\cite{Jian2018} this precise critical point on the triangular lattice would be described by a $N_f =4$ QED$_3$ with an emergent $PSU(4)=SU(4)/Z_4$ symmetry.
As a consequence of this enhanced symmetry, an exact degeneracy between the singlet and triplet monopoles in the spectrum should be observed. Indeed, we find the energy levels of the two types of monopoles close to being degenerate at the critical point $J_2/J_1=0.09$, while a larger splitting between these levels is observed throughout the remaining paramagnetic regime. In the case of an extended DSL region, the same degeneracy between singlet and triplet monopoles would be expected, while our results show growing energy splitting between them as $J_2/J_1\gtrsim 0.1$.

A second scenario is that the paramagnetic regime indeed realizes a stable Dirac spin liquid phase. Even though we do not observe the expected degeneracy between the two monopole excitations to be realized, we cannot rule out that the remaining energy splitting is still a finite-size effect. Recent work is concerned with the stability of a DSL as a function of $N_f$~\cite{Karthik2016,DiPietro2017,Albayrak2022}, and indications exist for the DSL to be a stable phase on the triangular lattice~\cite{He2022}. Empirical signatures of bilinear and monopole excitations had also been reported in a previous DMRG study~\cite{Hu2019}.

The data obtained from our exact diagonalization on small clusters does not allow to unambiguously distinguish between these two scenarios. From our data, both of the above scenarios are equally plausible. However, we think that evidence of either critical or long-range lozenge VBS correlations constitutes a new insight into the physics of the intermediate paramagnetic regime. We would like to point out, that a recent study has found evidence for a similar VBS state to be stabilized upon coupling the triangular lattice to phonons~\cite{Seifert2023}.

Finally, from previous work, we also notice that a chiral spin liquid (CSL) is in close vicinity to the paramagnetic regime when adding small scalar chirality interactions on the triangles of the form $\bm{S}_i\cdot (\bm{S}_j \times \bm{S}_k)$~\cite{Wietek2017}. A DSL can be unstable towards a magnetically ordered phase or a CSL phase through a QED3-Gross-Neveu quantum critical point~\cite{Dupuis2021}. While we do not observe direct evidence of a gapped CSL being realized, such a transition could be in close vicinity to the paramagnetic regime. Another point to clarify is whether the fact that the lowest parton ground state is reached for the centered boundary conditions with its complex (i.e.~time-reversal symmetry breaking) nature, indicates an instability towards a spontaneous chiral spin liquid formation. This effect would not appear in other geometries, such as the square torus studied in Ref.~\cite{Thomson2017}, where the time-reversal invariant antiperiodic boundary conditions minimize the energy.

While in our study we directly compared ansatz wave functions with eigenstates from ED on a finite cluster, we would like to point out that this technique could also be applied in complementary numerical techniques, for example, to construct excited states in tensor network simulations~\cite{Jin2020,Wu2020,Petrica2021} or other variational Monte Carlo techniques. This would allow also these techniques to study the non-trivial monopole and bilinear excitations revealed in the above on cylinder geometries or larger periodic tori. An application of the technology developed in this work to the kagome lattice $S=1/2$ Heisenberg model is the natural next step in the quest for quantum spin liquids.

\begin{acknowledgements}
We thank A. Vishwanath, S.~Sachdev, F.~Becca, J.~Motruk, M. Zaletel, and S. Bhattacharjee for stimulating discussions. We acknowledge PRACE for granting access to HPC resources at TGCC/CEA under grant number 2019204846SC. SC also acknowledges the use of HPC resources from CALMIP (grants 2021-P0677 and 2022-P0677) and GENCI (grant x2021050225). The Flatiron Institute is a division of the Simons Foundation. A.W. acknowledges financial support by the DFG through the Emmy Noether programme (509755282). AML acknowledges support by the Austrian Science Fund
(FWF) through project I-4548.
\end{acknowledgements}

\appendix

\section{Computation of overlaps using the Lanczos algorithm}
\label{app:overlaps}

Overlaps of ansatz wave functions with eigenstates of the Hamiltonian are a key analysis in the main text. Here, we summarize the technique proposed in Ref.~\cite{Gagliano1987} adapted to precisely the computation of overlaps. Let 
\begin{equation}
  \label{eq:tmatrix}
  T_n =
  \begin{pmatrix}
    \alpha_1 & \beta_1  & 0        &       & \cdots & 0 \\
    \beta_1  & \alpha_2 & \beta_2  & 0      &   & \vdots  \\
    0        & \beta_2  & \ddots   &        &        & \\
    &          &          &        & \ddots & 0 \\
    \vdots   &          &          & \ddots & \alpha_{n-1} & \beta_{n-1}\\

    0   &  \cdots &   & 0       & \beta_{n-1}       & \alpha_n \\
                     
  \end{pmatrix}.
\end{equation}
be the tridiagonal matrix of the Lanczos algorithm and,
\begin{equation}
  \label{eq:lanczosmatrixdef}
  V_n = \left( v_1 | \cdots | v_{n} \right ),
\end{equation}
the set of orthogonal Lanczos vectors $\ket{v_i}$. 
Let $\varepsilon_{k, n}$ and $\lambda_{k,n}$ denote the $k$-th eigenvalue and eigenvector of $T_n$, i.e. 
\begin{equation}
    T_n\lambda_{k,n} = \varepsilon_{k,n} \lambda_{k,n}.
\end{equation}
The eigenvalues $\varepsilon_{k, n}$ are conventionally referred to as Ritz values whereas the Ritz vectors are given by,
\begin{equation}
    \ket{\lambda_{k, n}} = \sum_{i=1}^n \lambda_{k, n}^{(i)} \ket{v_i},
\end{equation}
where $\lambda_{k, n}^{(i)}$ denotes the $i$-th entry of the vector $\lambda_{k, n}$. For extremal eigenvalues enumerated by $k$, we have
\begin{equation}
\ket{\lambda_{k, n}} \longrightarrow \ket{\lambda_{k}} \text{ for } n\rightarrow \infty,
\end{equation}
provided the usual prerequisite of convergence in the Lanczos algorithm with precise orthogonality of the Lanczos vectors is fulfilled. 

We are interested in computing the overlap of an eigenstate $|\lambda_k\rangle$ with a model wave function $|\phi\rangle$, i.e.
\begin{align}
\begin{split}
\label{eq:lanczosoverlaps}
|\braket{\phi | \lambda_k} | ^2 = \lim\limits_{n\rightarrow \infty} 
|\braket{\phi | \lambda_{k,n}} | ^2 = \lim\limits_{n\rightarrow \infty} |\sum_{i=1}^n \lambda_{k, n}^{(i)} \langle \phi | v_i\rangle | ^2
\end{split}
\end{align}
Now, if we choose the initial state of the Lanczos algorithm to be $\ket{v_1} \equiv \ket{\phi}$, we have $\braket{\phi | v_i} = \delta_{i,1}$ due to the orthogonality of the Lanczos vectors and \cref{eq:lanczosoverlaps} simplifies to,
\begin{equation}
\label{eq:lanczosoverlapssimple}
|\braket{\phi | \lambda_k} | ^2 = |\lambda_{k, n}^{(1)}|^2.
\end{equation}
\cref{eq:lanczosoverlapssimple} is used in all our overlap calculations in the main text. We check for convergence by analyzing the overlaps as a function of the Krylov space dimension $n$. The algorithm allows for computing overlaps with all Ritz vectors $\ket{\lambda_{k,n}}$ in a single Lanczos iteration without having to compute individual eigenstates. 

\section{Comparison of low-energy spectrum for the $N=36$  and $N=48$ clusters}
\label{app:spectrum_36_48}

\begin{figure}[t]
    \centering
    \includegraphics[width=\columnwidth]{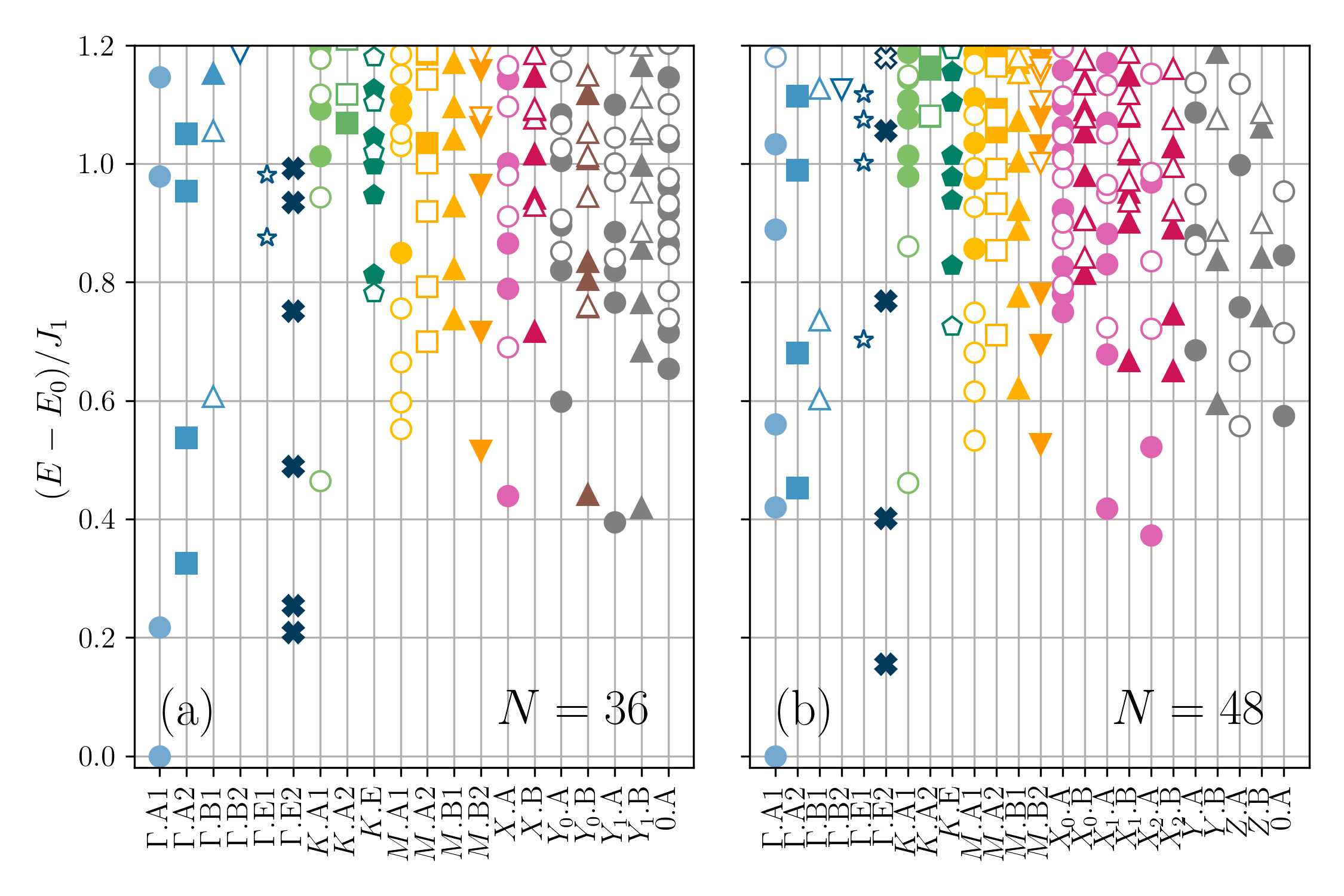}
    \caption{Comparison of the low-lying energy spectrum of the $J_1$-$J_2$ model at $J_2 / J_1 = 0.125$ on the (a) $N=36$ site cluster and the (b) $N=48$ site cluster.}
    \label{fig:j1j2_spectra_kresolved}
\end{figure}

To compare the energy spectra of the $J_1$-$J_2$ Heisenberg model obtained on the finite size clusters for $N=36$ and $N=48$, we show the momentum and point group symmetry resolved spectra in \cref{fig:j1j2_spectra_kresolved} for $J_2/J_1=1/8$. We observe that many low-lying energy levels have corresponding quantum numbers in both cases. The lowest excitation above the $\Gamma$.A1 ground state has a quantum number $\Gamma$.E2. Moreover, we observe low-lying singlet energy levels at the momentum in the exact middle between $\Gamma$ and $K$ ($X$.A for $N=36$ and $X_1$.A for $N=48$) and the $M$ point. Both these levels belong to the tower-of-states of the 12-site VBS, cf.~\cref{tab:andersontowermagtri}. Alternatively, these should become gapless in a DSL phase, scaling as $1/L$, where $L$ denotes the linear system size. Lastly, also a triplet level at $K$.A1 is observed at low energy in both cases, which in the language of QED$_3$ has been identified as the triplet monopole and is continuously connected to the first tower-of-states level of the $120^\circ$ N\'eel order. While the overall structure of the spectra appears to be in good agreement, it is unclear whether the levels scale according to either the VBS or the DSL scenario. For the VBS the splitting of the ground state degeneracy would be expected to be exponentially small in the system size, while a scaling of $1/L$ would be expected for the gapless points in the DSL phase. We would like to remark, that there can be non-trivial cluster-shape effects since the simulation cell spanning vectors of the $N=36$ and $N=48$ sites are not just multiples of one another.

\section{Dispersion of the single-particle states $\pi$-flux ansatz}
\label{app:dispersion}
In the following, we express the explicit formulae to construct the single-particle wave functions of the Dirac spin liquid used to construct the Gutzwiller projected wave functions in the main text. The elementary distance between neighboring lattice sites of the triangular lattice is set to $a=1$. As the $\pi$-flux ansatz is described by a two-site unit cell the free fermion Hamiltonian features two bands. Its momentum-dependent Hamiltonian is given by
\begin{widetext}
\begin{equation}
    \label{eq:twobandhamiltonian}
    H(k) = 2t
    \begin{pmatrix}
    \cos (\frac{k_x}{2}+ \frac{\sqrt{3} k_y }{2}) & - \cos(k_x) -i\sin\left(\frac{k_x}{2} - \frac{\sqrt{3}k_y}{2}\right) \\
    - \cos(k_x) + i\sin\left(\frac{k_x}{2} - \frac{\sqrt{3}k_y}{2}\right) & -\cos \left(\frac{k_x}{2} + \frac{\sqrt{3} k_y}{2}\right)
    \end{pmatrix}
\end{equation}
\end{widetext}

The dispersion of the two bands of the $\pi$-flux ansatz (i.e. the eigenvalues of) \cref{eq:twobandhamiltonian} is given by
\begin{equation}
\label{eq:diracdispersion}
    \varepsilon(k_x, k_y) = \pm t \beta(k_x, k_y)
\end{equation}
where 
\begin{align*}
\beta&(k_x, k_y) = \sqrt{2} \cdot\\
&\cdot \sqrt{ 3 + \cos(2k_x) - \cos(k_x - \sqrt{3} k_y) + \cos(k_x + \sqrt{3} k_y)}.
\end{align*}
The two Dirac nodes (i.e. zero modes in \cref{eq:diracdispersion}) are located at momenta 
\begin{equation}
    \bm{D}_{\pm} = \pm \frac{\pi}{2}(1, 1/\sqrt{3}).
\end{equation}
Taylor-expanding around the Dirac nodes gives a linear (relativistic) and rotationally invariant dispersion,
\begin{equation}
    E(\bm{k} - \bm{D}_\pm) \approx \sqrt{6} t ||\bm{k} - \bm{D}_\pm||
\end{equation}
Thus, the Fermi velocity is given by $v_F=\sqrt{6}t$. Using centered boundary conditions (see the Brillouin zone in Fig.~\ref{fig:overview}f), the minimal absolute momentum on a finite-size lattice $L\times L$ is located at a distance
\begin{equation}
    \parallel \bm{k} - D_{\pm} \parallel = \frac{4\pi}{3L}.
\end{equation}
Such boundary conditions are optimal for the energy of the filled Dirac sea before projection. 

The complex eigenvectors $v_1$ and $v_2$ of \cref{eq:twobandhamiltonian} are given by
\begin{align}
\label{eq:2bevecs}
\begin{split}
v^1_{\bm{k}} &= \frac{1}{\mathcal{N}}\left( -\alpha(k_x, k_y) - \beta(k_x, k_y), \gamma(k_x, k_y)\right)^T,  \\ 
v^2_{\bm{k}} &= \frac{1}{\mathcal{N}}\left( -\alpha(k_x, k_y) + \beta(k_x, k_y), \gamma(k_x, k_y)\right)^T,  
\end{split}
\end{align}
where $\bm{k}= (k_x, k_y)$, 
\begin{align}
\alpha(k_x, k_y) &= 2\cos\left(\frac{k_x}{2} + \frac{\sqrt{3}k_y}{2}\right), \\ 
\gamma(k_x, k_y) &= 2\cos(k_x) -2i\sin\left(\frac{k_x}{2} - \frac{\sqrt{3}k_y}{2}\right), 
\end{align}
and $\mathcal{N}$ denotes the real normalization constant normalizing the eigenvectors to unit norm. While in complex Hermitian matrices, the phase of the eigenvectors can be chosen arbitrarily, we adhere to the convention as in \cref{eq:2bevecs}, where the first component is chosen to be real. The precise form of the single-particle Bloch wave functions we are using is given by 
\begin{equation}
    \psi^b_{\bm{k}} (\bm{x}) = \textrm{e}^{i \bm{k}\cdot \bm{x}}
    v^b_{\bm{k}}(\bm{x}),
\end{equation}
where $b=0,1$ denotes the band index, and $v^b_{\bm{k}}(\bm{x})$ is either the first and second component of $v^b_{\bm{k}}$ in \cref{eq:2bevecs} depending on which one of the two sublattices the coordinate $\bm{x}$ is located on. Using this phase convention for the single-particle levels of the parton Ansatz, the phases of the Gutzwiller projected many-body states are uniquely determined.

\section{Tower-of-states analysis}
\label{app:tos}
Particular orders manifest themselves by a tower-of-states (TOS) structure of the energy spectrum on a finite-size lattice. A detailed description of how to derive the irreps in the TOS is given in Ref.~\cite{Wietek2017a}. \Cref{tab:andersontowermagtri} summarizes the predictions for orders discussed in the main text.
\begin{table}[h]
  \centering
  \begin{tabular}{|l|cccc|cccc|}
    \hline
    &\multicolumn{4}{l|}{$120^\circ$ N\'{e}el} &\multicolumn{4}{l|}{stripy order} \\
    \hline\hline
    $S$ & $\Gamma$.A1 & $\Gamma$.B1 & $K$.A1  & \quad & $\Gamma$.A1 & $\Gamma$.E2 & $M$.A  & \quad \\
    \hline
    0   &  1  & 0 & 0 & \quad & 1 & 1 & 0 & \quad \\
    1   &  0  & 1 & 1 & \quad & 0 & 0 & 1 & \quad \\
    2   &  1  & 0 & 2 & \quad & 1 & 1 & 0 & \quad \\
    3   &  1  & 2 & 2 & \quad & 0 & 0 & 1 & \quad \\
    \hline
    \hline
        &\multicolumn{4}{l|}{chiral spin liquid} &\multicolumn{4}{l|}{lozenge valence bond solid} \\
        \hline
        0 & \multicolumn{4}{l|}{$\Gamma$.A1 ($\times 2$),  $\Gamma$.E2} & \multicolumn{4}{l|}{$\Gamma$.A1, $\Gamma$.B2 , $K$.E , $M$.A1, $M$.B1, $X$.A, $X$.B}\\
        \hline
  \end{tabular}
  \caption{Multiplicities of irreducible representations (irreps) in the
    Anderson tower-of-states for the $120^\circ$ N\'eel and stripy magnetic orders as well as the valence bond solid and chiral spin liquid state on the
    triangular lattice. The irreps are labeled with total spin $S$, momentum, and point group irrep.}
  \label{tab:andersontowermagtri}
\end{table}

\section{Centered boundary conditions}
\label{app:centered_bv}
\begin{figure}
    \centering
    \includegraphics[width=\columnwidth]{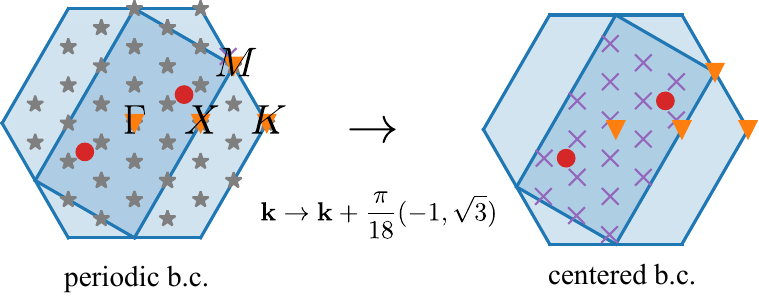}
    \caption{Comparison between momentum resolution with periodic and centered boundary conditions for the $N=36$ site simulation cluster.
    In the centered boundary conditions, the position of the Dirac nodes shown as red dots is at maximal distance from all resolved momenta.}
    \label{fig:centered_bc}
\end{figure}

Finite Bravais lattices with periodic boundary conditions define resolved momenta in reciprocal space. By twisting the boundary conditions, a physical flux through the incontractible loops of the torus can be introduced, which shifts the resolved momenta in reciprocal space. We find, that the kinetic energy of the Dirac spin liquid parton Hamiltonian is minimized whenever the resolved momenta are at a maximal distance from the Dirac nodes. This is achieved by shifting the resolved momenta by the particular momentum which "centers" the Dirac nodes. We refer to these boundary conditions as \textit{centered} boundary conditions. In particular, for the $N=36$ simulation cluster used in the main text, this is achieved by a shift $\bm{k} \rightarrow \bm{k} + \frac{\pi}{18}(-1, \sqrt{3})$. We show both the resolved momenta with periodic boundary conditions and centered boundary conditions in \cref{fig:centered_bc}. We note that the Dirac nodes on the $N=36$ site cluster can be centered either on the left-pointing triangles or the right-pointing triangles, which in principle yields two different ansatz wave functions. In the main manuscript, the centered boundary conditions with a shift vector, 
\begin{equation}
    \bm{k}_s = \frac{\pi}{18}(-1, \sqrt{3}), 
\end{equation}
have been used, centering the Dirac nodes on the left-pointing triangles. When using a square torus, the grid in momentum space is also a square lattice, and then the analogous shift vector procedure would result in antiperiodic boundary conditions, centering the Dirac nodes in a square plaquette in momentum space. 

\section{Construction of bilinear excitations}
\label{app:bilinear}
\begin{figure}
    \centering
    \includegraphics[width=\columnwidth]{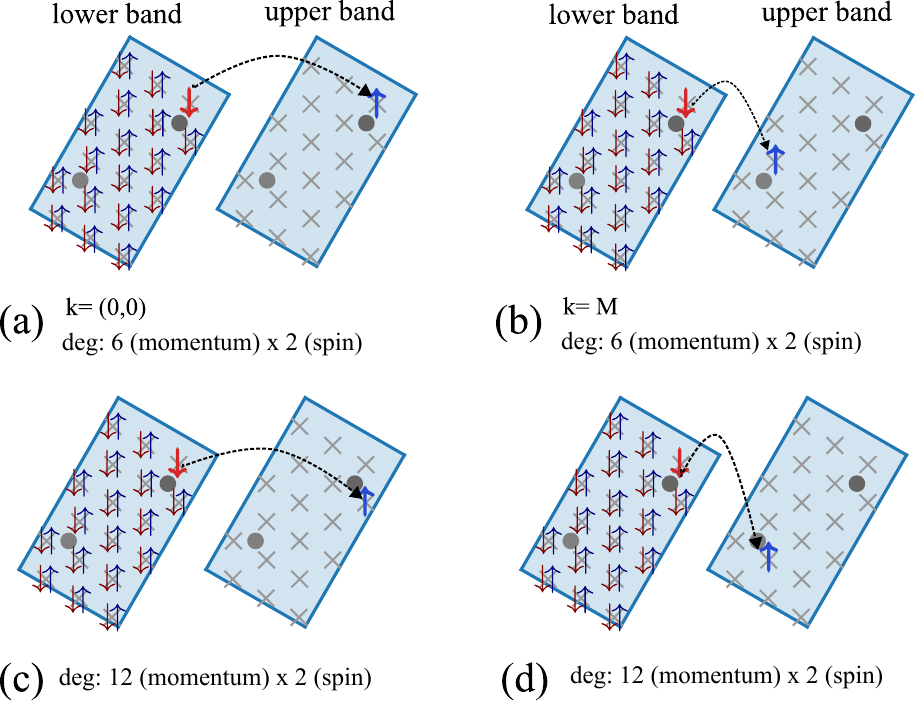}
    \caption{Bilinear excitations as particle-hole excitations of the parton ansatz with degeneracies. (a) excitation with $\bm{k}=0$ by exciting one parton to the upper band with exactly the same momentum. (b) excitation with $\bm{k}=M$ by exciting one parton to the corresponding momentum at the other Dirac valley. (c,d) excitation by moving one parton to another momentum either at the same (c) valley or the other valley(d).}
    \label{fig:bilinear_excitations}
\end{figure}

Bilinear excitations are constructed by performing particle-hole excitations out of the filled Dirac sea with centered boundary conditions. Gutzwiller projection conserves momentum and hence the momentum of these bilinear excitations can be derived from the momentum of the parton wave function. We only focus on excitations created from removing one parton from the lower band and inserting it into the upper band in one of the six degenerate energy levels with minimal energy. When exciting a parton from one momentum in the lower band to the same momentum in the upper band a net $\bm{k} = \Gamma = (0,0)$ state is obtained. If a parton is excited from a momentum in one valley to the corresponding momentum in the other valley, a net $\bm{k} = M$ state is obtained. Further excitations with momenta close to the $\bm{k}=\Gamma$ and $\bm{k}= M$ can be obtained by exciting a parton to the other four momenta. We summarize these possibilities in \cref{fig:bilinear_excitations}.

We notice that the total number of excitations with $S_z=0$ and momentum $\bm{k}=\Gamma$ is $12 = 6 \textrm{(momentum)} \times 2 \textrm{(spin)}$, cf. \cref{fig:bilinear_excitations}(a). Similarly, there are $12$ excitations with $S_z=0$ and momentum $\bm{k}=M$, cf. \cref{fig:bilinear_excitations}(b). Moreover, there are $48 = 6 \textrm{ (momentum } \bm{k} \textrm{)} \times 2 \textrm{(spin)} \times 4 \textrm{ (momentum } \bm{k^\prime \neq \bm{k}, \bm{k}+M} \textrm{)}$, cf. \cref{fig:bilinear_excitations}(c,d).
Besides these states, also excitations with total $S_z=\pm 1$ are possible, by exciting an $\uparrow$ parton (resp. $\downarrow$ parton) to a $\downarrow$ parton (resp. $\uparrow$ parton). This way another $72$ ansatz wave functions with $S_z=\pm1$ are constructed, which leads to a total number of $144$ possible bilinear excitations.

\section{Construction of monopole excitations}
\label{app:monopole}

\begin{figure}
    \centering
    \includegraphics[width=\columnwidth]{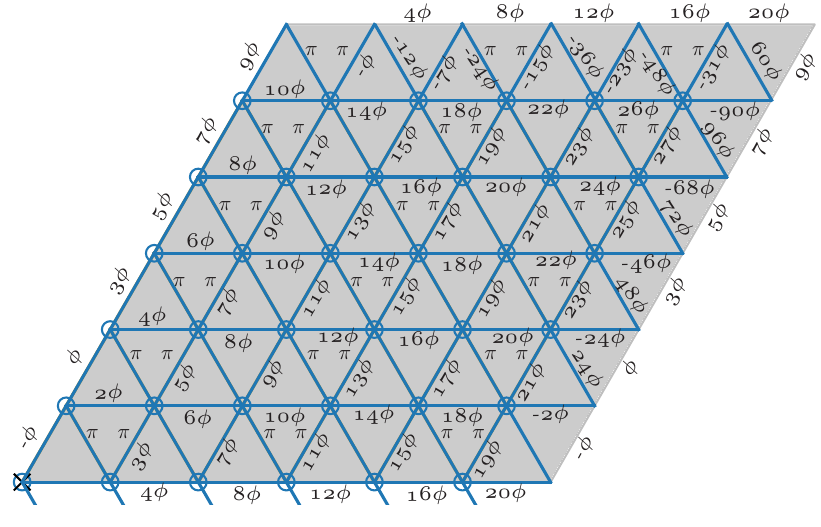}\\
    \includegraphics[width=\columnwidth]{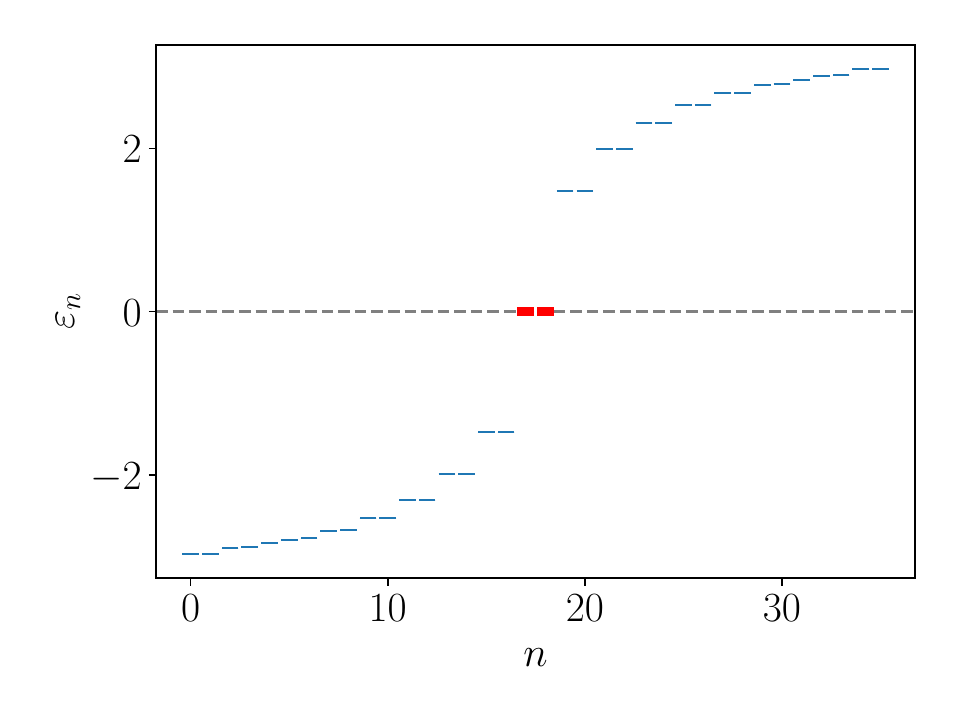}
    \caption{Top: flux pattern chosen to distribute one flux quantum over the $N=36$ triangular lattice so that all triangles contain the same additional flux $\Phi$ wrt the $\pi$ flux Ansatz. By convention, all fluxes point upwards or to the right. Bottom: corresponding tight-binding dispersion energies.}
    \label{fig:app_monopole}
\end{figure}

As sketched in Fig.~\ref{fig:overview}(c), putting a monopole inside the torus amounts to having a single flux quantum threading the whole system. By inspection, we have found one possible solution on the $N=36$ cluster that we consider, see Fig.~\ref{fig:app_monopole}(a) where we specify the phase factors appearing in the tight-binding model. Such a phase pattern provides a uniform flux of $\Phi=2\pi/(2N)$ per plaquette (on top of the $\pi$-flux Ansatz required for the DSL which amounts to having $\pi$ flux on all up triangles and zero flux on the down ones). The corresponding eigenspectrum of the tight-binding model is also given in Fig.~\ref{fig:app_monopole}(b), showing the existence of two zero-energy modes. Indeed, if the Dirac points are available for $\Phi=0$, then it is known that the monopole spectrum will exhibit $N_f$ degenerate states at zero energy (as found for instance in the quantum Hall effect of graphene).  Now in order to construct spin wave functions, we need to fill up all negative energy states and also put two fermions in the zero-energy modes, thus leading to 3 singlet and one triplet states~\cite{Song2020}. Similarly, we can consider "antimonopoles", i.e. a negative flux $\Phi=-2\pi/(2N)$ in order to get six additional states. In total, after Gutzwiller projection, we can generate six singlet and two triplet monopole wave functions. We have checked that they are linearly independent after projection.

\section{Construction and overlaps of two-monopoles}
\label{app:twomonopoles}
\begin{figure}
    \centering
    \includegraphics[width=\columnwidth]{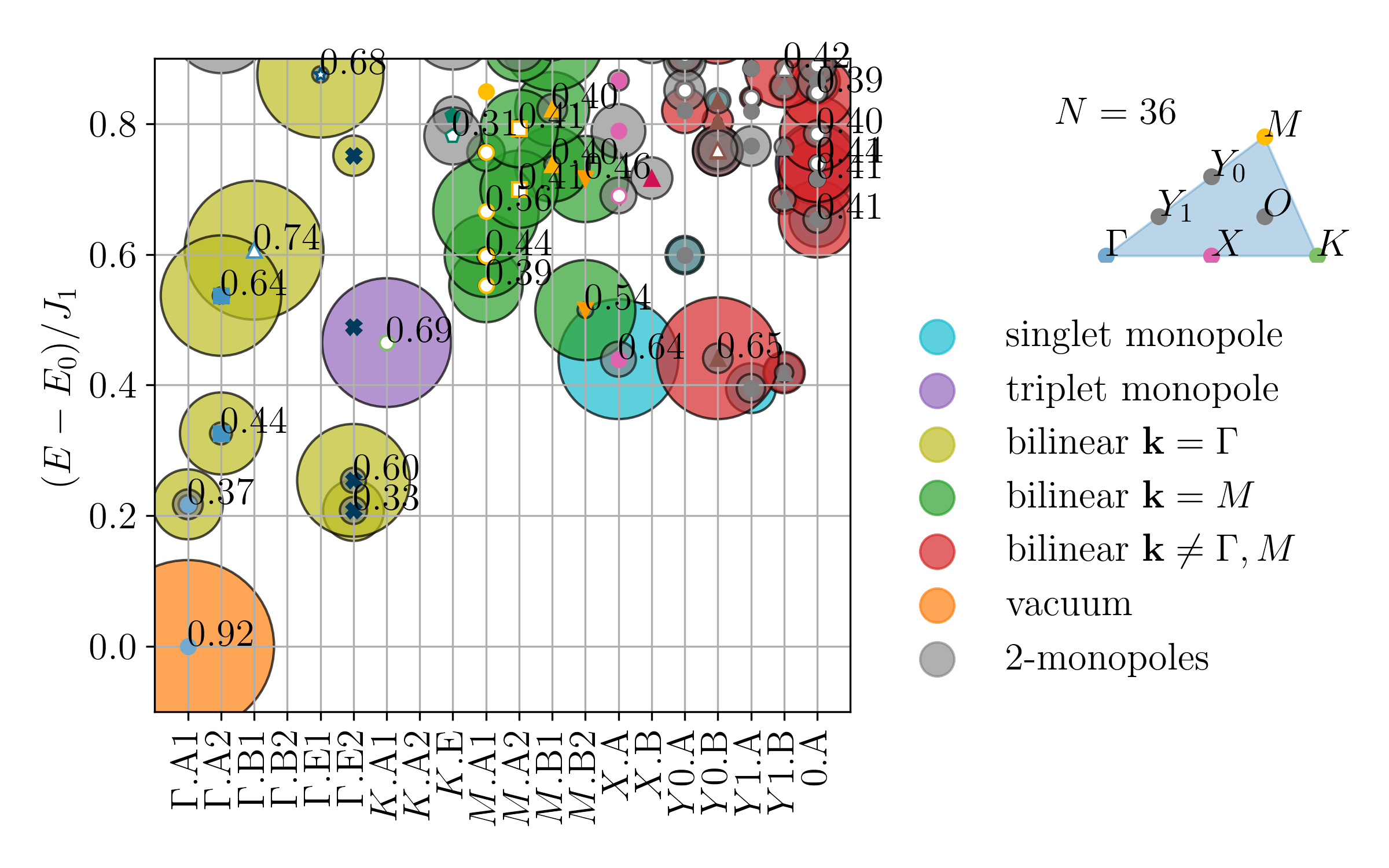}
    \caption{Overlaps $|\langle \psi | \psi_{ED} \rangle|$ of low-energy levels at $J_2/J_1 = 0.125$ with the vacuum state, singlet and triplet monopole states, as well as all bilinear excitations with $S_z=0$ and two-monopoles.
    The diameter of the colored circles is proportional to the overlap with the state at the center of the circle. We observe that almost every state in the dense low-energy spectrum has significant overlap with states constructed as elementary excitations of QED$_3$.}
    \label{fig:spectrumexhaustion_twomonopoles}
\end{figure}
In addition to the overlaps obtained from the single monopole excitations presented in the main text, we have also performed overlap calculations with ansatz wave functions for a two-monopole excitation at $J_2/J_1 = 0.125$. The results are shown in \cref{fig:spectrumexhaustion_twomonopoles}. These states are on average higher in energy than the single monopole and bilinear excitations. However, several low-lying states with energies $E<J_1$ still have significant overlap with these two-monopole excitations. The ansatz for these excitations is created analogously to the single-monopole excitations in \cref{fig:app_monopole}, where instead of having $2\pi $ flux through the system, we choose $4\pi$ flux.

Note that in such a situation, the tight-binding model exhibits 4 modes at approximately zero energy so that we can construct $2\times 6^2=72$ wave functions with a total $S_z=0$ for a positive or negative flux.

\begin{figure*}
    \centering
    \includegraphics[width=\textwidth]{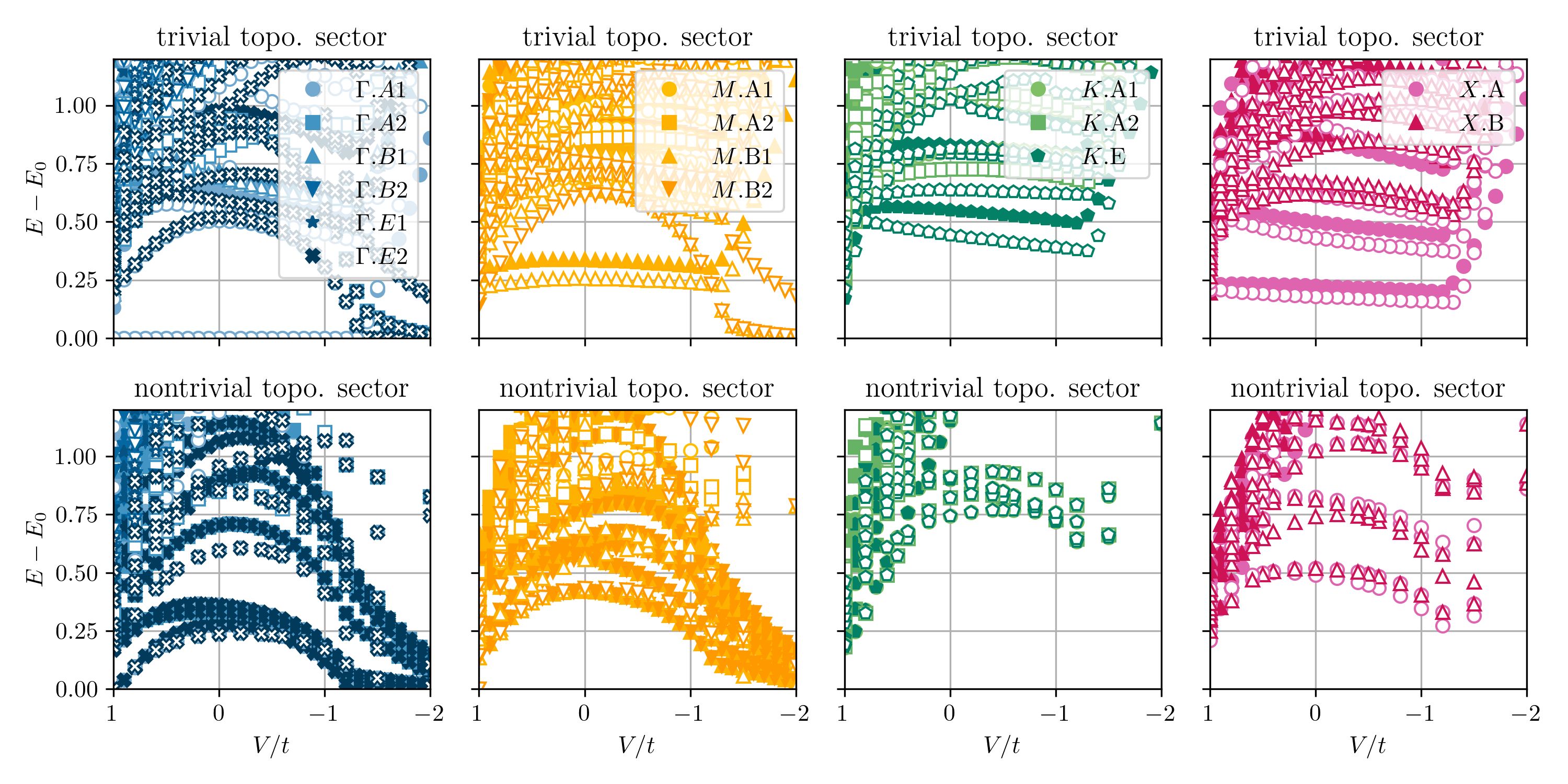}

    \caption{Excitation energies of the quantum dimer model at different momemta $\mathbf{k}= \Gamma, M, K, X$. The top row shows the spectra for the topologically trivial sector while the lower row shows the spectra for the other three topological sectors. Filled and empty symbols indicate the data from the $N=36$ and $N=48$ site clusters, respectively.}
    \label{fig:qdm_spectrum}
\end{figure*}

\section{QDM spectrum}

Let us consider the simplest quantum dimer model (QDM) on the triangular lattice~\cite{Rokhsar1988,Moessner2001}: 
\begin{equation}
\begin{split}
H_{\mathrm{QDM}}= & -t \sum_{{\includegraphics[scale=0.25]{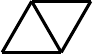}}} \left( | \raisebox{-0.65ex}{\includegraphics[scale=0.45]{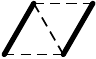}} \rangle  \langle \raisebox{-0.65ex}{\includegraphics[scale=0.45]{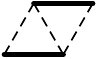}} | + | \raisebox{-0.65ex}{\includegraphics[scale=0.45]{dimer2}} \rangle  \langle \raisebox{-0.65ex}{\includegraphics[scale=0.45]{dimer1}} | \right) \\
& + V \sum_{{\includegraphics[scale=0.25]{plaq}}} \left( | \raisebox{-0.65ex}{\includegraphics[scale=0.45]{dimer1}} \rangle  \langle \raisebox{-0.65ex}{\includegraphics[scale=0.45]{dimer1}} | + | \raisebox{-0.65ex}{\includegraphics[scale=0.45]{dimer2}} \rangle  \langle \raisebox{-0.65ex}{\includegraphics[scale=0.45]{dimer2}} | \right).
\end{split}
\label{eq:QDM_triangular}
\end{equation}

As we argue, level spectroscopy is a useful tool to investigate low-lying excitations of quantum many-body systems. In Fig.~\ref{fig:qdm_spectrum}, we provide ED spectra obtained for the QDM on $N=36$ and $N=48$ clusters, the same clusters as used in Fig.~\ref{fig:j1j2_spectrum} for the spin-1/2 Heisenberg model. Note that for readability we plot the excitation energies as a function of decreasing $V/t$ starting from the Rokhsar-Kivelson point $V/t=1$ and we do not discuss the staggered phase found at $V/t>1$~\cite{Moessner2001}. The phase diagram is known to include a $Z_2$ topological phase ($1>V/t \gtrapprox 0.8$), a 12-site unit cell VBS ($0.8 \gtrapprox V/t \gtrapprox -0.75$) and a columnar phase beyond~\cite{Moessner2001,Ralko2005}. The presence of several low-lying $\Gamma.E2$ levels in the $J_1-J_2$ Heisenberg model motivates us to compare the spectrum with the QDM model at $V/t=-1$, where the QDM is close to the transition toward the columnar physics, a suggestive similarity with the Heisenberg model, which enters the directional, stripy magnetic phase for larger values of $J_2/J_1$.

\begin{figure}
    \centering
    \includegraphics[width=\columnwidth]{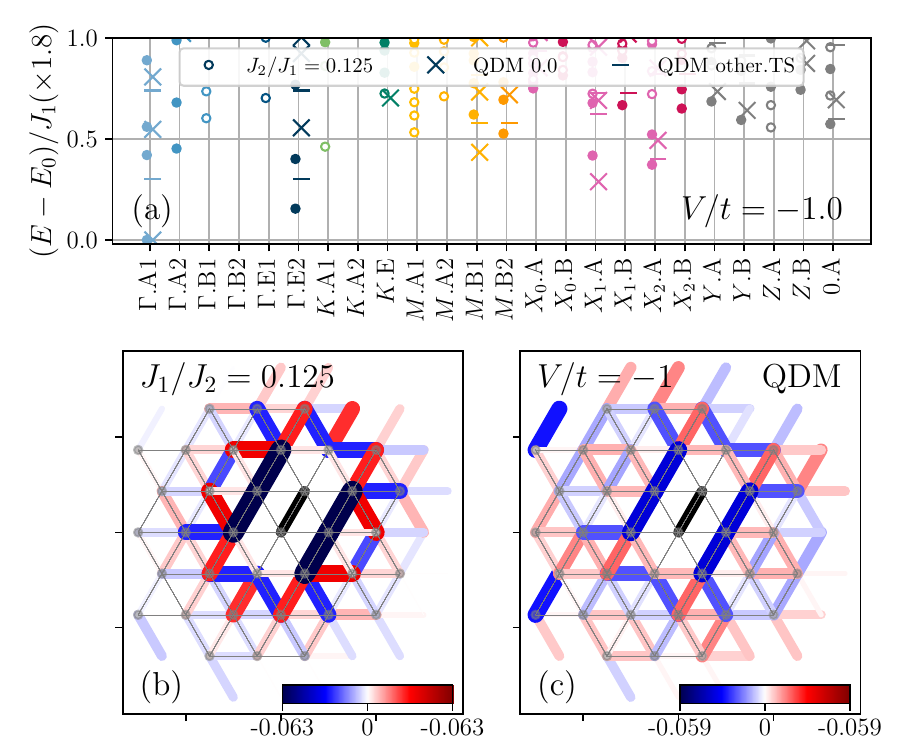}
    \caption{(a) Comparison of the low-lying energy spectrum of the $J_1$-$J_2$ model at $J_2 / J_1 = 0.125$ and the QDM model at $V/t=-1.0$ on the $N=36$ cluster. The filled (open) circles denote even (odd) spin levels.
    (b) Connected dimer correlations $\langle(\bm{S}_0\cdot \bm{S}_1)(\bm{S}_i \cdot \bm{S}_j)\rangle_c$ of the ground state from ED of the $J_1$-$J_2$ model at $J_2 / J_1 = 0.125$ on the $N=36$ cluster. (c) Connected dimer correlations of the QDM in the VBS phase at $V/t=-1.0$. This figure is the analog of \cref{fig:vbsqdm} with $N=36$ in the main text.}
    \label{fig:dimer_corr}
\end{figure}

\section{Dimer correlations for the $N=36$ cluster}
\label{app:dimer_corr}

For the QDM, a 12-site valence bond crystal was clearly established~\cite{Moessner2001,Ralko2005}. For comparison, we plot in Fig.~\ref{fig:dimer_corr} the dimer correlations obtained on \emph{the same} cluster of $N=36$ sites, both for the QDM in the VBS phase and for the Heisenberg model in the spin liquid one. We do observe a qualitative agreement in the patterns of these dimer correlations, compatible with the 12-site VBS at least on short length scales.

\bibliography{main}

\end{document}